\documentclass{epl}
\usepackage{graphicx}
\title{Power Laws, Precursors and Predictability During Failure}
\author{Rumi De\inst{1}\thanks{E-mail: \email{rumi@mrc.iisc.ernet.in}} \and G. Ananthakrishna\inst{1,2}\thanks{E-mail: \email{garani@mrc.iisc.ernet.in}}}
\institute{
\inst{1} Material Research Center, Indian Institute of Science,
         Bangalore-560012, India.\\
\inst{2} Center for Condensed Matter Theory, Indian Institute of Science,
Bangalore-560012, India.
       }
\pacs{74.25.Ld}{Mechanical and acoustical properties, elasticity and ultrasonic
attenuation}
\pacs{91.30.Px}{Phenomena related to earthquake prediction}
\pacs{64.60.Ht}{Dynamic critical phenomena}

\begin{document}

\maketitle

\begin{abstract}
We investigate the dynamics of a modified Burridge-Knopoff model
by introducing a dissipative term to mimic the bursts of acoustic
emission (AE) from rock samples. The model explains many features
of the statistics of AE signals observed in experiments such as
the crossover in the exponent value from relatively small
amplitude AE signals to larger regime, and their dependence on the
pulling speed. Significantly, we find that the cumulative energy
dissipated identified with acoustic emission can be used to
predict a major slip event. We also find a data collapse of the
acoustic activity for several major slip events describable  by a
universal stretched exponential with corrections in terms of
time-to-failure.
\end{abstract}

Predicting failure of materials  is of interest   in science and
engineering (electrical breakdown, fracture of laboratory samples
to engineering  structures) particularly so in seismology due to
the enormous damage earthquakes can cause.  Whether it is at a
laboratory  or geological scale, this amounts to identifying
useful precursors at a statistically significant level. One
important non-destructive tool in fracture studies is the acoustic
emission (AE) technique as it is sensitive to the microstructural
changes occurring in the sample. Insight into earthquake dynamics
has been obtained through  fracture studies of ( usually precut
samples to mimic slip on preexisting tectonic faults)  rock
samples \cite{Lockner91}. Such studies have established that there
is a considerable overlap between AE and seismology as both are
concerned about the generation and propagation of elastic waves.
Quite early, the statistics of the AE signals was shown to exhibit
a power law \cite{Mogi62,Lockner91,Scholz68b} similar to the
Gutenberg-Richters law for the magnitudes of earthquakes
\cite{Gutenberg} and Omori's law for aftershocks
\cite{Mogi62,Main96}. These prompted further investigations to
look for precursor effects that can be used for earthquake
predictability \cite{Main96,Sam92,Main89,Hainzl}.

Apart from the power law distributions  observed in the AE signals
during fracture, acoustic activity of unusually large number of
situations as varied as volcanic activity \cite{Diodati},
micro-fracturing process \cite{Petri94,Hans95}, and collective
dislocation motion \cite{Miguel}, exhibits power laws statistics.
Though the general mechanism attributed to AE is the release of
stored strain energy, the details are system specific. Thus, the
ubiquity of the power law statistics of AE signals suggests that
the details of the underlying processes are irrelevant. Indeed,
the conceptual framework of self-organized criticality (SOC) was
introduced to explain the universality of power law statistics in
varied systems \cite{Bak,Jensen}. This brings up a related {\it
puzzle} that  lack of intrinsic length scales and time scales in
power law systems  has been taken to mean {\it lack of
predictability of individual  avalanches } \cite{Main96,Debate99}.
This generally held opinion has also triggered considerable debate
regarding earthquake predictability, as seismically active fault
systems are considered to be in a SOC state \cite{Debate99}. Thus,
it would be desirable to design  a model which exhibits power law
statistics where predictability of individual events is possible.

On the other hand, there has been reports of increased levels of
AE signals before failure of rock samples
\cite{Scholz68a,Lockner96,Sam92}, in experiments on laboratory
samples \cite{Guari02,Johan} and precursory effects in individual
earthquakes as well \cite{Main96,Main89,Sykes90,Bufe93}. There has
been attempts to predict failure within the framework of
time-to-failure (TTF) models, both on the laboratory level
\cite{Guari02} as well on geological scale
\cite{Huang98}. (Some efforts have also been made
on the predictability of avalanches in SOC state \cite{Rosen}.)
However, as the primary aim of TTF models is to mimic the rate of
increase in precursory variable, they are independent of the
nature of the variable. Thus, if one is interested explaining the
origin of  increasing level of activity in AE signals observed in
experiments as the system approaches failure,  it is necessary to
first model AE in terms of displacement related variable as AE
corresponds to high frequency elastic waves. Here, we model AE
signals by introducing an additional dissipative term into
Burridge-Knopoff model \cite{BK,Carlson,Schmit} that helps us  to
identify precursory effect and hence to predict slip events.

Our second objective is to explain some recent results on  AE
studies on rock samples. These studies  report an  interesting
crossover in the exponent value from small amplitude regime to
large \cite{Yabe}, a result that is similar to the well noted
observation on the change in the power law exponent for small and
large magnitudes earthquakes \cite{Scholz92}. The exponent value
is also found to be sensitive to the deformation rate
\cite{Yabegrl}. To the best of our knowledge, {\it there has been
no explanation of these observations}. This can partly be traced
to the lack of efforts to model AE signals in terms of {\it
displacement related variables}.

Deformation and/or breaking of the asperities results in an
accelerated motion of the local areas of slip where the stored
potential energy is converted to kinetic energy. The general
mechanism attributed to acoustic emission is the sudden release of
the stored strain energy. In the case of modeling AE signals
involving abrupt dislocation motion, the energy of AE signals,
$E_{ae}(r)$ is taken to be proportional $\dot \epsilon^2(r)$,
where $\dot \epsilon$ is the local plastic strain rate
\cite{Weiss}. However, in general there is spatial inhomogeneity
which implies that total energy $E_{ae}\propto \int (\nabla \dot
\epsilon)^2 d^3r$.  For  the BK  model, displacement rate takes
the role of $\dot \epsilon$.  On the other hand, $E_{ae}$, has the
same form as the Rayleigh dissipation functional \cite{Land}
arising from the rapid movement of a localized region. Such a
rapid movement prevents the system from attaining a quasi-static
equilibrium which in turn generates dissipative forces that resist
the rapid motion.  On general grounds this dissipative forces  has
been shown to be proportional to $\int (\nabla v)^2 d^3 r$, where
$v $ is the local velocity. Such a term is termed as solid
viscosity in parallel with shear viscosity of fluids. (We ignore
other types of dissipation such as radiation damping.) Indeed, we
have found that the above term mimics all the features of the AE
signals observed during martensitic transformation \cite{Rajeev} (
both the power law statistics of the AE signals during  thermal
cycling and  the highly correlated AE signals associated with
growth and shrinkage of martensite domains).

\begin{figure}[!h]

\twofigures[height=3.5cm,width=7.0cm]{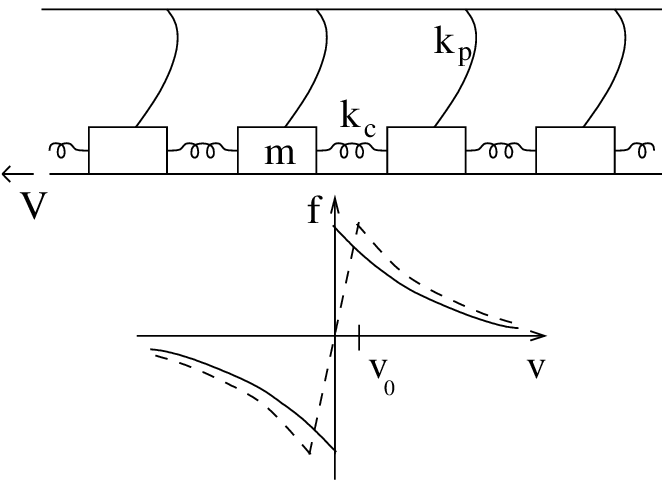}{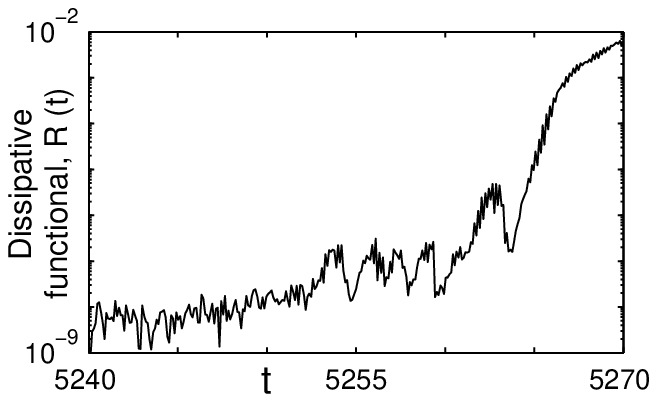}
\caption{The Burridge-Knopoff spring block model and the two forms
of friction laws used.} \label{bkmodel} \caption{ Dissipative
functional, R(t) as a function of t.}\label{Rt}
\end{figure}

The Burridge-Knopoff (BK) model for earthquakes \cite{BK} and its
variants are popular models among the physics community. Despite
its limitation (lack of appropriate continuum limit)
\cite{Schmit,Rice}, it forms a convenient platform to investigate
the question of predictability of large avalanches as the
dissipated energy bursts (in our modified model) themselves follow
a power law distribution. The model consists of a chain of blocks
of mass $m$ coupled to each other by coil springs of strength
$k_c$ and attached to a fixed surface by leaf springs of strength
$ k_p$ as shown in Fig. \ref{bkmodel}. The blocks are in contact
with a rough surface moving at constant speed $v$ (mimicking the
points of contact between two tectonic plates). A crucial input
into the model is the velocity-dependent frictional force between
the blocks and the surface (see Fig. \ref{bkmodel}).

The additional dissipative force  is introduced through the
Lagrange's equations of motion given by ${d\over dt}({{\delta
L}\over{\delta \dot u(x)}}) - {{\delta L}\over{\delta u(x)}} = -
{{\delta F}\over{\delta\dot u(x)}}$ with the Lagrangian $L = T -
P$. The kinetic energy $T$ is defined by $T =
{m\over{2}}\int({{\partial u(x)}\over{\partial t}})^2dx$,   and
the potential energy by $P = {1\over2}\int[k_c({{\partial
u(x)}\over{\partial x}})^2 + k_p({ u(x)})^2]dx$.  Here $ u$ is the
displacement of the blocks measured from the initial equilibrium
position. The Rayleigh dissipative functional \cite{Land} is given
by $R = {\gamma\over{2}}\int({{\partial\dot u(x)}\over{\partial
x}})^2dx$, where $\gamma$ is a dissipation coefficient. The total
dissipation F is the sum of $R(t)$ and frictional dissipation
given by $F_{fr} = \int[f(v + \dot u(x))]dx$, where the over dot
refers to the time derivative and the frictional force is taken to
be derivable from a potential like function. Then the equation of
motion is
\begin {equation}
m\frac{\partial^2 u}{\partial t^2} =
k_c\frac{\partial^2u}{\partial x^2} - k_pu - \frac {\partial
f(\dot u+v)}{\partial \dot u} + \gamma \frac{\partial^2\dot
u}{\partial x^2},
\end {equation}
The discretized version, in the notation of Ref \cite{Carlson},
reads
\begin {equation}
\ddot U_j = l^2(U_{j+1} - 2U_j + U_{j-1} ) - U_j - \phi (2\alpha
\nu + 2\alpha \dot U_j)
 + {\gamma_c} ( \dot U_{j+1} -2\dot U_j + \dot U_{j-1} ),
\end {equation}
\noindent where $ U_j$ is the dimensionless displacement of the
$j^{th}$ block, $\nu $ the dimensionless pulling velocity,$l^2 =
k_c/k_p$ the ratio of the slipping time to the loading time, and
$\alpha$ is the rate of velocity-weakening in the scaled
frictional force $\phi$ (see Fig. \ref{bkmodel}). The Coulomb
frictional law is shown by the solid line. The dashed curve uses a
resistive creep branch ending at $v_{\circ}$ ($\sim 10^{-7}$ here)
beyond which the velocity weakening law operates. $\gamma_c$ is
the scaled dissipation coefficient. The continuum limit of Eq. (2)
exists for the creep branch (even in the absence of $R(t)$) which
ensures a length scale below which all perturbations are damped
\cite{Carlson}. Such a length scale is absent for the Coulomb case
even as the continuum limit exists due to the additional
dissipative term. These differences  between the frictional laws
may influence the nature of the statistics which we investigate in
detail.

This model without the last term has been extensively studied
\cite{BK,Carlson,Schmit}. Starting from random initial conditions
for all the blocks, slip events ranging from one-block event to
those extending over the entire fault (occurring roughly once in a
loading period $\tau_L \sim 2/\nu$ ) are seen in the steady state
for both types of frictional laws.

Equation (2) has been solved  using a fourth-order Runge-Kutta
method with open boundary condition for both types of frictional
laws shown in Fig. \ref{bkmodel}.  Random initial conditions are
imposed. After discarding the initial transients, long data sets
are recorded when the system has reached a stationary state. The
parameters used here are $l=10, \alpha = 2.5, N = 100, 200$ for
$\nu = 0.01$ and 0.001 and a range of values of $\gamma_c$. The
modified BK model produces the same statistics of slip events as
that without the last term in Eq. (2) as long as the value of
$\gamma_c$ is small, typically $\gamma_c < 0.5$. The results
presented here are for $N=100$ and $\gamma_c = 0.02$.

Since the rate of energy dissipated \cite{Land} due to local
accelerating blocks is given by $dE_{ae}/dt = -2R(t)$, we
calculate $R(t)$ which exhibits  bursts similar to the AE signals
observed in experiments. A plot of $R(t)$ for the case with
Coulomb frictional  law is shown in Fig. \ref{Rt}. For the case
with creep branch, $R(t)$ is less noisy.

\begin{figure}[!t]
\twofigures[height=4.5cm,width=7.0cm]{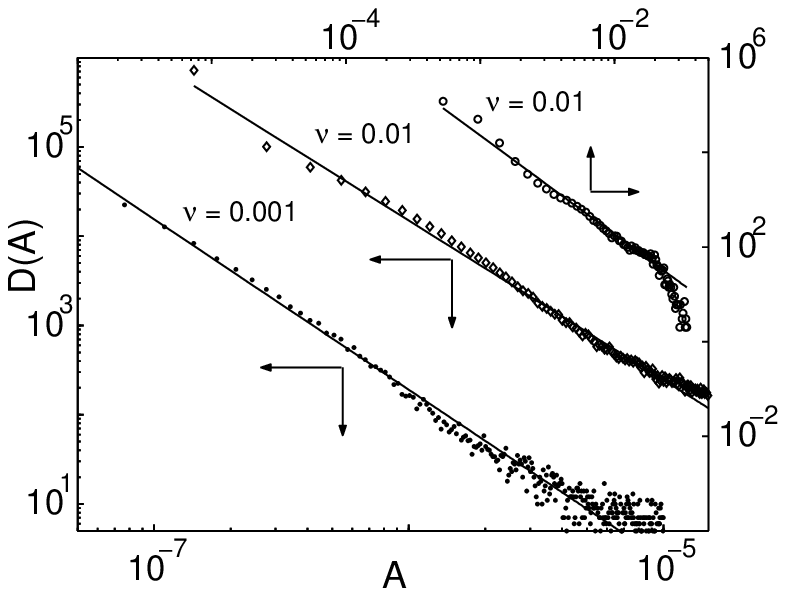}{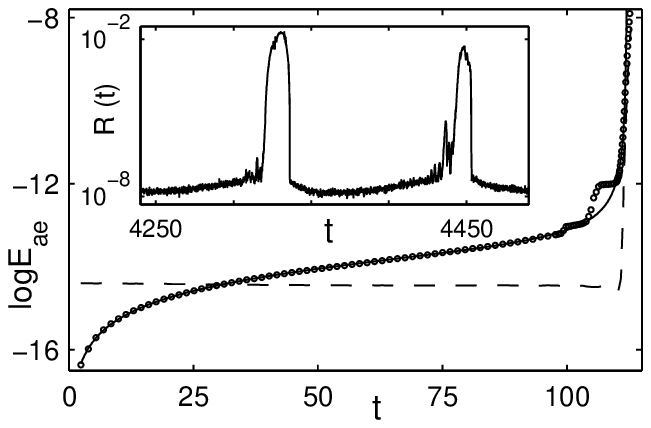}
\caption{Distribution of amplitudes of $R(A)$, $D(A)$ versus A.
for small amplitudes ($\diamondsuit$ for $\nu =0.01$ $\bullet$ for
$\nu = 0.001$ ) and large amplitudes (o for $\nu =0.01$ shifted up
for clarity).} \label{amppow} \caption{A semi-log plot of the
cumulative energy, $E_{ae}$ versus t ($\circ$) along with the fit
(solid line). Dashed line corresponds to the mean kinetic energy.
The inset shows R(t) as a function of t. } \label{Rfit}
\end{figure}

We now consider the statistics of the energy bursts $R(t)$.
Denoting $A$ to be the amplitude of $R(t)$ ({\it i.e.}, from a
maximum to the next minimum), we find that the distribution of the
magnitudes $D(A)$ (accumulated over $2 \times 10^5$ time steps),
shows a power law  $D(A)\sim A^{-m}$. Instead of a single power
law anticipated, we find that the distribution shows two regions,
one for relatively smaller amplitudes and another for large values
shown by the two distinct plots for the Coulomb case shown in Fig.
\ref{amppow}. ( The crossover in the power law exponent from
smaller to larger amplitude region is seen as a scatter at the
upper end of $A \sim 10^{-5}$.)  Similar results are obtained when
the frictional law has a creep branch. The value of $m$ for the
small amplitudes region ( typically  $ < 10^{-4}$) is $\sim 1.78
\pm 0.01$, significantly smaller than that for large amplitudes
which is $\sim 2.09 \pm 0.02$, consistent with the recent
experimental result on AE of rock samples \cite{Yabe}.  The
increase in the exponent value mimics a similar observation for
large magnitude earthquakes ( $> 7.0$ on the Richter scale
\cite{Scholz92}).

Recently  Yabe {\it et al.} \cite{Yabegrl} have reported that the
exponent value corresponding to relatively small amplitude regime
increases with decreasing deformation rate while that for the
large amplitude regime is found to be insensitive. To check this,
we have calculated $D(A)$ for $\nu =0.001$ shown in Fig.
\ref{amppow} which shows that the exponent for small amplitude
increases to $ 1.91 \pm 0.02$. However, we find  that the exponent
for larger amplitude regime is insensitive (not shown) to the
changes in $\nu$.

This result can be physically explained by
analyzing the influence of the pulling velocity on slip events of
varying sizes. We first note that $R(t)$ depends on the difference
in the velocities of neighboring blocks. The velocity of
`microscopic' events (small number of blocks) has been shown to
proportional to $\nu$ \cite{Carlson}. For single block events, as
the neighboring blocks are at rest, the number of such events are
fewer in proportion to the pulling speed, both of which are
evident from Fig. \ref{amppow}. For the two block events, the
contribution comes mostly from the edges as the difference in the
velocities of the two blocks are of similar magnitude. In a
similar way, it can be argued that for slip events of finite size,
the extent of the contribution to $R(t)$ is decided by the
ruggedness of the velocity profile within the slipping region; the
magnitude of $R(t)$ is lower if the velocity is smoother. The
ruggedness of the velocity profile, however, is itself decided by
how much time the system gets to `relax'. At lower values of
$\nu$, there is sufficient time for the blocks to attain nearly
the same velocities as the neighboring blocks compared to that at
higher $\nu$ values. Thus, the larger slip events contribute
lesser to $R(t)$ for smaller $\nu$ values and hence  the slope of
$log D(A)$ increases for smaller value of $\nu$.

Now we consider the possibility of a precursor effect. A plot of
$R(t)$ is shown in the inset of Fig. \ref{Rfit} for the frictional
law with a creep branch.  {\it A gradual increase in the activity
of the energy dissipated can be seen which accelerates just prior
to the occurrence of a `major' slip event.} The rapid increase in
$R(t)$ coincides with the abrupt increase in the mean kinetic
energy (KE). Here, we have used KE as a measure of event size  as
it is a good indicator of the magnitude of the slip events. We
find similar increase in $R(t)$ for all `major' slip events. (In
our simulations, the KE of observable events ranges from $10^{-4}$
to 0.2. We refer to all such events as `major' events.) This
suggests that $R(t)$ can be used as a precursor for the onset of a
major slip event. As $R(t)$ is noisy, a better quantity for the
analysis is the cumulative energy dissipated $E_{ae}(t)$ ($\propto
\int_0^t R( t^{\prime} ) d t^{\prime} $). $E_{ae}(t)$ grows in a
stepped manner with their magnitudes increasing as we approach a
major event. A plot of $log E_{ae}$ is shown in Fig. \ref{Rfit}
along with a fit (continuous curve) having the functional form,
\begin{equation}
log E_{ae}(t) =  -a_1 t^{-\alpha_1} [ 1 - a_2 \vert (t-t_c)/t_c
\vert ^{-\alpha_2}].
\end{equation}
Here, $t$ is the time measured from some initial point after a
major event. The constants $a_1,a_2,\alpha_1$ and $\alpha_2$ and
$t_c$ are adjustable. The crucial parameter $t_c$ is the time of
occurrence of a major slip event (often referred to as the
`failure' point). It is clear that the fit is striking. Given a
reasonable stretch of the data, the initial increasing trend in
$log E_{ae}$ is easily fitted to a stretched exponential,{\it
i.e.}, $- a_1 t^{-\alpha_1}$. The additional term is introduced to
account for the observed rapid increase in the activity as we
approach the major event. In contrast, the mean kinetic energy
increases abruptly on reaching the  event  ( dashed line). It is
clear that the estimated $t_c$ agrees quite well with that of KE
of the event.

\begin{figure}[!t]
\twofigures[height=4.0cm,width=7.0cm]{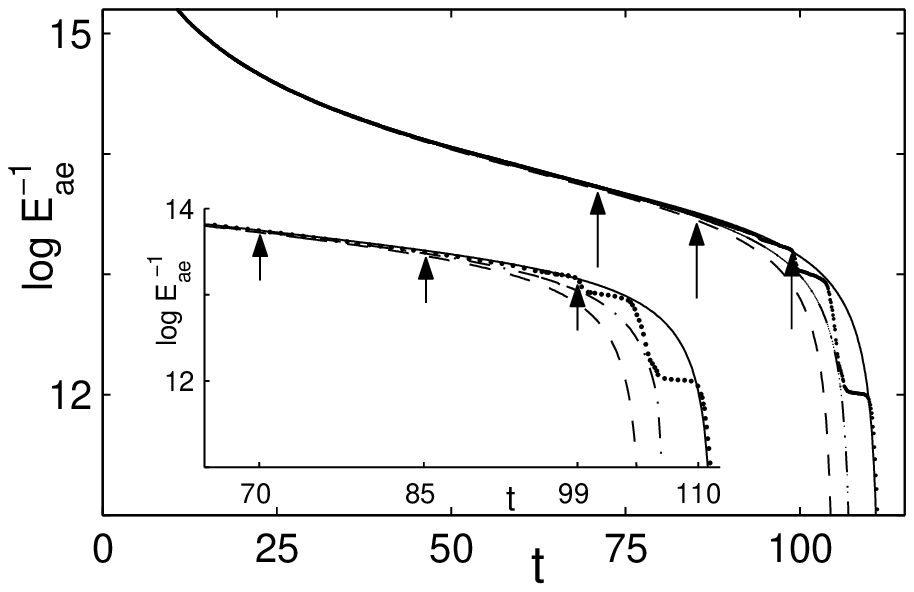}{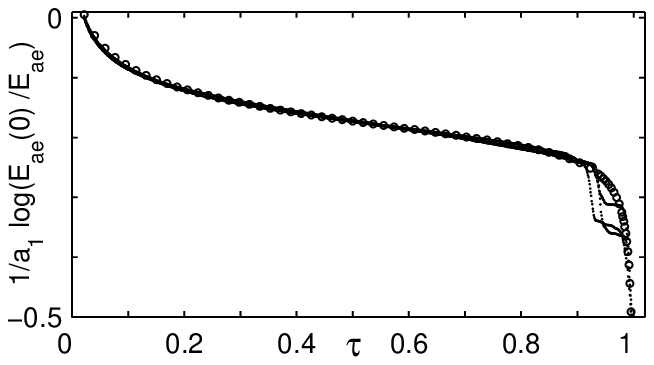}
\caption{A  plot of  $log E_{ae}^{-1}$ versus t. Inset shows the
enlarged section at time $t_1$ ($-$), $t_2$ ($-\cdot$) and $t_3$
(solid line). Data shown ($\cdot$) is indistinguishable from the
fit till $t_3$.} \label{invR} \caption{Collapsed data using $
a_1^{-1} log (E_{ae}(0)/E_{ae}(\tau))$ vs. $\tau$ for three
different events along with the fit shown by ($\circ$).}
\label{univr}
\end{figure}

Now, we address the question of predictability of major slip
events using $E_{ae}(t)$. This is equivalent to determining the
correct $t_c$.  Given $E_{ae}(t)$ over a reasonable initial
stretch of time, say till $t_1$ (the first arrow in Fig.
\ref{invR}), we find that the four constants $a_1,a_2,\alpha_1$
and $\alpha_2$ are already well determined (within a small error
bar). These change very little with time. (Only $t_c$ changes.) A
fit to Eq. (3) also gives $t_c^{(1)}$ at $t_1$ ( which can only be
considered as an estimate based on the data till $t_1$). One such
curve is shown by a dashed line with the arrow shown at $t_1$.
However, as time progresses, the data accumulated later usually
deviates from the predicted curve if $t_c$ is inaccurate as is the
case for the fits till $t_1$ and $t_2$  for instance (Fig.
\ref{invR}). If on the other hand, the deviation of the predicted
curve from the accumulated data decreases with passage of time
within the error bar (as is the case for the region just before
$t_3$),  then, the value of $t_c$ is likely to be accurate.
Indeed, the extrapolated continuous curve corresponding to data
fit till $t=t_3$ (third arrow in Fig. \ref{invR}) with the
predicted $t_c^{(3)}$ is seen to follow the data very well.
(Usually, the data deviates from the predicted curve with a sudden
decrease in $E_{ae}^{-1}$ which is again an indication of a
coherent slipping of several blocks before the onset of a fully
delocalized event. But the general trend soon follows the
extrapolated curve.) Then, $t_3$ can be taken as the warning time
for the onset of the major event. The actual $t_c$ read off from
the kinetic energy plots is 111.0 where as the predicted $t_c$  is
111.6 giving the accuracy in the prediction of correct $t_c$ to be
$\sim 99.5\%$.  This fit is obtained when $t_{last}$ is 12\% away
from the true $t_c$. Similar  numbers are obtained for several
other events  fitted. If the approach to all slip events is
described by the scale invariant form, then one should expect to
find a data collapse for different events. Indeed, in terms of a
scaled time $\tau = t/t_c$, we find that the data corresponding to
different events collapses into a single curve given by
\begin{equation}
 a_1^{-1} log[E_{ae}(0)/E_{ae}(\tau)] = \tau^{-\alpha_1} [1 -a_2 \vert (\tau
-1)\vert ^{-\alpha_2}] + a_1^{-1} logE_{ae}(0) .
\end{equation}

A plot corresponding to  three different events is shown in Fig. \ref{univr}
along with the fit. The results are similar when the Coulomb frictional
law is used except that $R(t)$ is more noisy and hence prone to
slightly larger errors in predicted $t_c$.

In summary, the model captures several experimentally observed
features of  acoustic emission such as the two different  exponent
values in the power law statistics  for small and large amplitude
regimes with the former being more sensitive to the pulling speed.
The dependence of the exponent on the pulling speed has been
traced to the form of $R(t)$, namely, the gradient of the local
velocity. More significantly, the analysis shows that it is
possible to predict a major event fairly accurately. This result
demonstrates that {\it power law statistics in itself does not
preclude predictability} (of individual events). Further, we note
that a SOC state demands that all observable quantities should
follow a scale invariant form which is {\it clearly respected} by
Eqs. (3,4) representing the approach to all events. The data
collapse for different events into a scale free form clearly
suggests that the dynamics of approach to major events is
universal.

A few comments are in order. We stress that this precursor effect
is absent in the total kinetic energy or seismic moments. As there
is no correspondence between seismic moments or  the KE of the
events with $R(t)$ (which depends on the difference between the
velocities of neighboring elements), our analysis cannot predict
the magnitudes of the slip events. Indeed, in some cases we find
that $R(t)$ is larger for a smaller slip event, as is the case for
the two peaks in $R(t)$ shown in Fig. \ref{Rfit}.  We point out
here that the gradual increase in the energy dissipated as we
approach a major event  is different from that reported in the
context of the BK model \cite{Shaw}. Our work also differs from
the approach of Huang {\it et al.} \cite{Huang98} in the sense
that in their analysis, the hierarchical structure and long range
interaction are necessary ingredients for the power law approach
to failure with log periodic corrections that arises due to
discrete scale invariance \cite{Johan,Huang98}. As far as we know,
this is the first time that reproduces several unexplained
experimental results on AE of rock samples \cite{Yabe,Yabegrl}. We
believe that it would be possible to address the problem of
predictability in other power situations, which  however, may
demand  capturing the essential physics of the relevant precursor
effect.

RD wishes to thank Dr. S. Rajesh and Dr. M. S. Bharathi for usefull
discussions. This work is supported by Department of Science and Technology, Grant No. Sp/s2K-26/98, New Delhi, India.

\end{document}